\documentstyle[12pt]{article}
 
\begin{document}


\newcommand{\be}[1]{\begin{equation}\label{#1}}
\newcommand{\beq}{\begin{equation}}
\newcommand{\ee}{\end{equation}}
\newcommand{\beqn}[1]{\begin{eqnarray}\label{#1}}
\newcommand{\eeqn}{\end{eqnarray}}
\newcommand{\bd}{\begin{displaymath}}
\newcommand{\ed}{\end{displaymath}}
\newcommand{\mat}[4]{\left(\begin{array}{cc}{#1}&{#2}\\{#3}&{#4}\end{array}
\right)}
\newcommand{\matr}[9]{\left(\begin{array}{ccc}{#1}&{#2}&{#3}\\
{#4}&{#5}&{#6}\\{#7}&{#8}&{#9}\end{array}\right)}
\def\simlt{\mathrel{\lower2.5pt\vbox{\lineskip=0pt\baselineskip=0pt
           \hbox{$<$}\hbox{$\sim$}}}}
\def\simgt{\mathrel{\lower2.5pt\vbox{\lineskip=0pt\baselineskip=0pt
           \hbox{$>$}\hbox{$\sim$}}}}
\def\unity{{\hbox{1\kern-.8mm l}}}
\def\tanb{\tan\beta}
\def\epr{E^\prime}
\def\al{\alpha}
\def\ga{\gamma}
\def\Ga{\Gamma}
\def\om{\omega}
\def\OM{\Omega}
\def\la{\lambda}
\def\La{\Lambda}
\newcommand{\eps}{\varepsilon}
\def\ep{\epsilon}
\newcommand{\ov}{\overline}
\renewcommand{\to}{\rightarrow}
\def\mcirc{{\stackrel{o}{m}}}
\newcommand{\bM}{\bar M} 
\newcommand{\cM}{{\cal M}} 
\newcommand{\cO}{{\cal O}} 
%

\begin{flushright}
hep-ph/9611277 ~~~~~ INFN-FE 13/96 \\ 
October 1996 \\
\end{flushright}
\vspace{10mm}

\begin{center}
{\Large \bf Anomalous U(1) Symmetry and \\
Missing Doublet SU(5) Model } 
\end{center} 

\vspace{0.3cm}
\centerline{\large Zurab Berezhiani ~ and ~ Zurab Tavartkiladze }
\vspace{4mm} 
\centerline{\it INFN Sezione di Ferrara, 44100 Ferrara, Italy,}
\centerline{and} 
\centerline{\it Institute of Physics, Georgian Academy of Sciences, 
380077 Tbilisi, Georgia} 
\vspace{1.9cm}

\begin{abstract} 
We present the supersymmetric $SU(5)$ models which provide a simple 
``all order'' solution to the doublet-triplet splitting 
problem through the missing doublet mechanism.
The crucial role is played by the anomalous $U(1)_A$ gauge symmetry 
and no additional discrete or global symmetries are needed. 
Remarkably, such models can be realized even if the 75-plet Higgs 
is replaced by the standard 24-plet. 
The same $U(1)_A$ symmetry can also guarantee an exact or approximate 
conservation of R parity, by suppressing the B and L violating 
operators to the needed level. The neutrino masses and the proton decay 
via $d=5$ operators are also examined. We also extend the model 
by incorporating $U(1)_A$ as a horizontal symmetry for 
explaining the fermion mass and mixing hierarchy. 
Interestingly, in this scheme the necessary mild violation of the 
troublesome $SU(5)$ degeneracy between the 
down quark and the charged lepton masses 
can be induced by certain R-parity violating operators. 
\end{abstract}

\newpage

\section{Introduction} 

Many realistic string models contain the gauge $U(1)$ factors 
with non-vanishing traces over the charges of the matter 
superfields. One can find their linear combination $U(1)_A$ 
which can be `truly' anomalous while the other combinations are 
rendered traceless. 
Existence of such an anomalous $U(1)_A$ symmetry does not imply 
an anomaly in the `progenitor' string theory. 
In the field theory limit it can be understood as a result of 
truncating the string spectrum to the particle spectrum, 
and all mixed anomalies of the matter fields are effectively canceled 
via the ``universal'' Green-Schwarz mechanism \cite{GS}. 
As it was shown \cite{DSW}, the D-term of such $U(1)_A$ symmetry 
gets a non-zero Fayet-Iliopoulos term $\xi$ related to the 
string scale $M_{\rm str}=gM_P$ as: 
\be{Dterm}
D_A = \xi + \sum Q_i |\varphi_i|^2 , ~~~~~ 
\xi= \frac{M_{\rm str}^2}{192\pi^2} {\rm Tr}Q
\ee 
where the sum includes all scalar fields $\varphi_i$ present 
in the theory with the nonzero $U(1)_A$ charges $Q_i$, 
and the relevant mass scale is the (reduced) Planck scale 
$M_P=(8\pi G_N)^{-1/2}\simeq 2\cdot 10^{18}$ GeV.
As a result, some of the scalar fields which VEVs would vanish 
in the absence of the Fayet-Iliopoulos term $\xi$ now can get 
the non-zero vacuum expectation value (VEV) $\sim \sqrt{\xi}$. 
Therefore, the $U(1)_A$ symmetry breaking scale 
is naturally small but not too small -- the ratio 
\be{xi}
\frac{\sqrt{\xi}}{M_P} \sim 10^{-1} - 10^{-2} 
\ee
is just of order of the fermion mass ratios 
in the neighbouring families. Following this observation, 
in the literature anomalous gauge symmetry $U(1)$ is widely used 
as a horizontal symmetry for explaining the fermion mass 
hierarchy \cite{IR}. 
Recently the anomalous $U(1)_A$ symmetry was applied \cite{Gia} 
to the solution of the doublet-triplet splitting problem by means 
the Goldstone boson mechanism in the supersymmetric 
$SU(6)$ theory \cite{su6}.  
This is perhaps the most economic and transperent way to get 
an accidental global symmetry $SU(6)\times SU(6)$ in the Higgs 
superpotential, among the other proposals which make use of the 
discrete symmetries \cite{BDM}. 

In the present paper we show that the idea of the anomalous 
$U(1)_A$ gauge symmetry inspired by the string theory can be 
used also for achieving a simple `all order' solution 
to the doublet-triplet splitting problem via the 
missing doublet mechanism (MDM) \cite{MDM}  
in the supersymmetric $SU(5)$ theory. In particular, in sect. 2 
we reproduce the original MDM by arangement of the $U(1)_A$ 
charges of the Higgs 5- and 50-plets, which solution is stable 
against the Planck scale corrections. In the sect. 3 
we suggest the improved missing doublet models 
which can reconcile between the proton lifetime and 
perturbative regime of the $SU(5)$ gauge constant above 
the GUT scale. Then we study the implications of this model 
for the neutrino masses (sect. 4), naturall suppression of the 
dimension 3 and 4 B and L violating (R parity violating) operators 
(sect. 5), Planck scale induced dimension 5 B and L violating 
operators (sect. 6) and then present a model incorporating 
also the fermion mass picture (sect. 7). 
Finally, in sect. 8 we briefly discuss our results.   


\section{The missing doublet SU(5)$\times$U(1)$_A$ model } 

Consider the $SU(5)$ model with the Higgs sector 
containing superfields in the following representations: 
$H\sim 5$, $\bar H\sim \bar5$, $\Phi\sim 75$, $\Psi \sim 50$,   
$\ov{\Psi}\sim \ov{50}$, and a singlet $X$, and three families 
of the fermion superfields $10_i+\bar5_i$, $i=1,2,3$.\footnote{
Usually the Higgs 
and fermion superfields are distinguished by introducing 
the matter parity: positive for Higgses and negative for fermions. 
Below we show that in our model we need not to introduce 
the {\em ad hoc} matter parity and it can emerge as an 
automatic consequence of the anomalous 
$U(1)_A$ charges. }   
Let us assume that $X$ has a negative $U(1)_A$ charge, $Q_X=-q<0$, 
while the other charges are arranged as follows:  
$Q_\Phi=0$, $Q_{\ov{\Psi}}=-Q_H=-h$, $Q_{\Psi}=-Q_{\bar H}=q+h$, 
$Q_{10}=-\frac{1}{2}h$ and  $Q_{\bar5}= q+\frac{3}{2}h$.  
Then the most general renormalizable Higgs superpotential reads as 
\beqn{W75}
& W_{\rm Higgs} = W_1 + W_2 ,  \nonumber \\ 
& W_1 = M \Phi^2 + \la\Phi^3 , \nonumber \\  
& W_2 = \la_1 X\Psi\ov{\Psi} +\la_2 H\Phi\ov{\Psi} + 
\la_3 \bar H\Phi\Psi,  
\eeqn
where the order one constants are understood in the trilinear terms, 
and the mass parameter $M$ is of order GUT scale 
$M_G\simeq 10^{16}$ GeV. 
Note, that both the mass term $\mu H\bar H$ and the coupling 
$XH\bar H$ are forbidden by the $U(1)_A$ symmetry. 

The couplings of $H,\bar H$ to the fermion fields 
$\bar5$ and 10 are the following: 
\be{Yukawa}
W_{\rm Yuk}=
\Ga^u_{ij} 10_i 10_j H  + \Ga^d_{ij} 10_i \bar5_j \bar H 
\ee
where $\Ga^{u,d}_{ij}$, $i,j=1,2,3$ are the Yukawa coupling constants. 

One has to analyze the superpotential (\ref{W75}) together with the 
D-terms 
\be{D} 
g_5^2 \left(\sum \phi_r^\dagger T_a^{(r)} \phi_r \right)^2 + 
g_A^2 \left(\sum Q_r|\phi_r|^2 - q|X|^2 + \xi \right)^2 ,  
\ee
where under $\phi_r$ we imply the scalar components of all 
superfields present in the theory besides $X$ with their $U(1)_A$ 
charges $Q_r$, $T_a^{(r)}$ are the $SU(5)$ generators in the proper 
representations $r$,  
and $g_5$ and $g_A$ respectively are the $SU(5)$ and $U(1)_A$ 
gauge constants. It is easy to see that there exists a 
supersymetry conserving vacuum when 75-plet $\Phi$ gets the 
VEV $\langle\Phi\rangle = \frac{M}{4\la}$ 
which breaks $SU(5)$ down to 
$SU(3)\times SU(2)\times U(1)$ symmetry of the MSSM.  
This VEV is normalized so that masses of the X-and Y-gauge 
`dinosaurus' are 
$M_G= 24g_5^2 V_{\Phi}^2$. 
As for the field $X$, its non-zero VEV 
$\langle X \rangle = \sqrt{\xi/q}$ emerges from the 
anomalous D-term in 
(\ref{D}) and it is essentially close to the GUT scale $M_G$: 
modulo the factor $({\rm Tr}Q/q)^{1/2} \sim 1-10$ we have 
$\langle X \rangle \sim 5\cdot 10^{16}$ GeV.\footnote{
In our model the scales $\langle \Phi\rangle$ 
and $\langle X\rangle$ are actually independent 
and their numerical closeness seems rather accidental.  
Needless to say that it would be highly interesting to have a 
realistic mechanism which would naturally explain both values 
from the same origin. } 
All other VEVs are vanishing, which is consistent with 
the usual F- and D-flatness conditions. 

For the numerical estimates in the following we take 
the following values for the $U(1)_A$ symmetry breaking 
scale: $\langle X\rangle=10^{17}$ GeV and 
$\langle X\rangle=4\cdot 10^{17}$ GeV. Correspondingly, 
for the ratio $\eps= \langle X\rangle/M_P$ we have 
$\eps\simeq 1/20$ and $\eps\simeq \sqrt{1/20}$. 
The first situation could emerge if $({\rm Tr}Q)$ has a 
moderate value in units of $q$. 
In the second one we need $({\rm Tr}Q)/q \sim 100$, which 
could be indeed the case in the models considered below. 
For the $SU(5)$ symmetry breaking scale we take a standard value 
$\langle\Phi\rangle=2\cdot 10^{16}$ GeV, though in view of 
the various threshold corrections it could be somewhat larger 
or smaller.    


After substituting the VEVs $\langle \Phi \rangle$ and 
$\langle X \rangle$ in $W_2$, the 50-plets receive a mass 
$M_{\Psi}=\la_1 \langle X\rangle$  while the colour triplet 
components ($H_3,\bar{H}_3$)  in $H,\bar H$ get masses via 
the `seesaw' mixing to the triplets 
$\Psi_3,\ov{\Psi}_3$ from $\Psi,\ov{\Psi}$. 
The relevant mass matrix reads as  
\be{T}
\begin{array}{cc} & {\begin{array}{cc} 
\bar{H}_{3} & \,\,\,\;\ov{\Psi}_{3} \end{array}}\\ 
\vspace{2mm}
\cM_3 = \begin{array}{c}
H_3 \\ \Psi_3 \end{array}\!\!\!\!\!&{\left(\begin{array}{cc}
0 & \la_{2}\langle\Phi\rangle \\ \la_{3}\langle\Phi\rangle 
& M_{\Psi} \end{array}\right)} 
\end{array}  
\ee 
Hence, all triplet fields are massive. The triplets in 50-plets 
have a mass of order $\langle X\rangle >\langle\Phi\rangle$ 
while the ones contained 
in 5-plets get mass $M_T\sim \langle\Phi\rangle^2/\langle X\rangle$. 

As for the doublet components $H_2,\bar{H}_2$ in $H,\bar H$, 
they remain massless since the 50-plets do not have doublet 
fragments. Therefore, they can be identified with the 
MSSM Higgs doublets: $H_2=H_{u}$, $\bar{H}_2=H_{d}$.  
In this way, we have achieved a simple solution of the 
doublet-triplet splitting problem in the $SU(5)$ theory, 
in the spirit of the original MDM \cite{MDM}.  
Note however that our solution is stable against the 
Planck scale corrections 
since the higher order operators cutoff by $M_P$ 
\be{MP} 
\frac{X^k\Phi^n}{M_P^{k+n-1}}  H\bar H 
\ee 
are forbidden for any integer $k$ and $n$ 
by the $U(1)_A$ charges of the relevant fields.  
Indeed, the combination $H\bar H$ has a negative $U(1)_A$ charge 
$Q_{\bar{H}}+Q_H=-q$ and thus it cannot be 
compensated by the powers of $X$. 

The MDM however has a generic problem related to 
the baryon number violating $d=5$ operators \cite{dim5}. 
Indeed, if $M_\Psi \sim \langle X \rangle \geq 10^{17}$ GeV, 
then the triplets $H_3,\bar{H}_3$ are too light 
($M_T \leq 10^{15}$ GeV).  
In terms of the big matrix of the triplet masses (\ref{T}) 
the cutoff scale for the relevant $d=5$ operators is 
\be{dim5}
(\cM_3^{-1})_{11} = 
\frac{\la_1\langle X\rangle}{\la_2\la_3\langle\Phi\rangle^2} 
=\frac{1}{M_T}  
\ee 
and thus they mediate too fast proton decay \cite{Nath}, 
which is excluded by the present experimental data \cite{PDG}.

If $M_\Psi \leq M_G$,\footnote{This can be the case if 
$\la_1\ll 1$, or if 50-plets get mass from higher order operator, 
e.g. $\frac{X^2}{M_P} \Psi\ov{\Psi}$. } 
then the model becomes strongly interacting above the scale $M_G$ 
since now besides the 75-plet, also the 50- and 
$\ov{50}$-plets contribute the renormalization group 
running of the $SU(5)$ gauge constant $g_5$. This contributions 
would drive the running $g_5$ out of the perturbative regime 
below the string scale 

All these put the `canonical' version of the MDM \cite{MDM} 
in a ``no-go'' situation unless due to GUT scale 
threshold uncertainties or maybe at the price of 
introducing some additional states at intermediate scales  
the $SU(5)$ unification scale $M_G$ could rise up to 
about $10^{17}$ GeV \cite{Yamada}.   
In the next section we present an improved version of the 
missing doublet model which do not suffer from the above problems.


\section{Improved missing doublet model (IMDM) }

We employ the proposal of ref. \cite{HMTY} to introduce two 
sets of the 5- and 50-plets: 
$H,H'\sim 5$, $\bar{H},\bar{H}'\sim \bar5$, $\Psi,\Psi'\sim 50$ and 
$\ov{\Psi},\ov{\Psi}'\sim \ov{50}$. Let us prescribe the $U(1)_A$ charges 
to these fields as follows: 
\beqn{Q-new}
& Q_X=-q, ~~~~ Q_{\Phi}=0, \nonumber \\ 
& Q_H=h, ~~~ Q_{\bar H}=-(n+3)q-h, ~~~
 Q_{H'}=(n+2)q+h, ~~~ Q_{\bar H'}=-q-h,
\nonumber \\ 
& Q_{\Psi}=q+h, ~~~ Q_{\ov{\Psi}}=-h, ~~~
Q_{\Psi'}=(n+3)q+h, ~~~ Q_{\ov{\Psi}'}=-(n+2)q-h, 
\eeqn 
where $n=0,1,2,\dots$ is an integer number. 
Now the Yukawa couplings (\ref{Yukawa}) require the following 
$U(1)_A$ charges of the fermions:\footnote{We still prescribe to 
fermions family independent charges assuming the  
hierarchy of the Yukawa constants has an {\em ad hoc} origin, 
or it emerges as a result of the spontaneously broken 
(Abelian or non-Abelian) horizontal symmetry, 
in the spirit of refs. \cite{FN,su3H}. }    
\be{Q-f}
Q_{10}=-\frac12 h, ~~~~ Q_{\bar 5}= (n+3)q + \frac32 h  
\ee

Then the self-interaction terms $W_1$ of $\Phi$ in the Higgs 
superpotential are still the same as in (\ref{W75}) 
while the terms in $W_2$ are modified as: 
\be{W2-R} 
W_2 = \la_1 X\Psi\ov{\Psi} +\la'_1 X\Psi'\ov{\Psi}' + 
\la_2 H\Phi\ov{\Psi} + \la'_2 H'\Phi\ov{\Psi}' + 
\la_3 \bar H\Phi\Psi' + \la'_3 \bar{H}'\Phi\Psi 
\ee 
If only these couplings are left, then one can easily make sure 
that all triplet fields in the theory are massive, and 
there is no proton decay via their exchanges. However, the 
additional couple of the massless doublets $H'_2 + \bar{H}'_2$ 
emerges in the particle spectrum, which would spoil the 
$SU(5)$ unification of the gauge couplings. 

However, now we include also the non-renormalizable 
operators cutoff by the Planck scale:
\be{W2-NR}
W'_2 =  \al_1 \frac{X^{n+1}}{M_P^n} H'\bar{H}' + 
\al_2 \frac{X^{n+2}}{M_P^{n+2}}H'\Phi\ov{\Psi} +  
\al_3 \frac{X^{n+2}}{M_P^{n+2}}\bar{H}'\Phi\Psi' + 
\al_4 \frac{X^{n+3}}{M_P^{n+2}} \Psi'\ov{\Psi}   ,  
\ee 
(with respect to $\Phi$ only the lowest dimensional operators 
are shown). These operators could emerge after integrating out 
the heavy states with masses $\sim M_P$ \cite{FN}.  

After substituting the VEV $\langle X\rangle$, all these couplings 
together can be expressed in a matrix form as\footnote{
It was attempted in ref. \cite{HMTY} to obtain this pattern 
by using the Peccei-Quinn symmetry. However, these authors have 
omitted `by hands' the possible large mass terms $M_P H\bar{H}'$ 
and $M_P H'\bar{H}$ which were allowed by their $U(1)_{PQ}$ 
symmetry. } 
\beqn{MDM-new}
&\hspace{-6mm} 
\bar H ~~~~~~ \bar{H}' ~~~~~~ \ov{\Psi} ~~~~~~ \ov{\Psi}' &
\nonumber \\
\begin{array}{cccc}
H ~ \\
H' ~ \\
\Psi ~ \\
\Psi' ~ \\
\end{array}&
\hspace{-6mm}\left(
\begin{array}{cccc}
0 &  0    & \Phi  & 0 \\
0 & \eps^{n+1}M_P & \eps^{n+2}\Phi & \Phi  \\ 
0 & \Phi & \eps M_P & 0 \\
\Phi & \eps^{n+2}\Phi & \eps^{n+3}M_P  & \eps M_P  
\end{array} \right)&  
\eeqn
where $\eps = \langle X\rangle/M_P$, 
and the coupling constants are omitted. 
Note, that all zero elements in this expression are ``all order'' 
zeros in $X/M_P$, since the $U(1)_A$ charges of the corresponding 
terms are all negative. The entries $\sim \eps^{n+2(3)}$ 
are not relevant for our further estimates and we have shown them 
only for demonstrating their relatively small magnitudes.  


Therefore, after integrating out the heavy 50-plets at the scale 
$V_X$, we obtain the following mass matrices for the doublet and 
triplet fragments in 5-plet Higgses: 
\beqn{D-T}
& \begin{array}{cc} & {\begin{array}{cc} 
\bar{H}_2 & \,\,\,\;\bar{H}'_2 \end{array}}\\ 
\vspace{2mm}
\cM_2 = \begin{array}{c}
H_2 \\ H'_2 \end{array}\!\!\!\!\!&{\left(\begin{array}{cc}
0 & 0 \\ 0 & M_D \end{array}\right)} 
\end{array}  , ~~~~~ M_D\sim \eps^{n+1}M_P    \nonumber \\ 
%
& \begin{array}{cc} & {\begin{array}{cc} 
\bar{H}_3 & \,\,\,\;\bar{H}'_3 \end{array}}\\ 
\vspace{2mm}
\cM_3 = \begin{array}{c}
H_3 \\ H'_3 \end{array}\!\!\!\!\!&{\left(\begin{array}{cc}
0 & M_T \\ M'_T & M_D \end{array}\right)} 
\end{array}  , ~~~~~ 
M_T, M'_T \sim \langle\Phi\rangle^2/\langle X\rangle 
\sim \eps_G^2\eps^{-1} M_P 
\eeqn 
where $\eps_G=\langle\Phi\rangle/M_P\sim 10^{-2}$. 
Thus, $H_2$ and $\bar{H}_2$ remain massless and thus they can be 
identified to the MSSM Higgses $H_{u,d}$, while 
$H'_2,\bar{H}'_2$ get mass $M_D$. 
As for the triplet fragments, they acquire mass terms 
$M_T \sim \eps_G\eps^{-1} M_G$, i.e. below the GUT scale $M_G$. 
Nevertheless, the fact that $M_T < M_G$ now does not necessarily 
leads to unacceptably fast proton decay. In particular, if $M_D=0$ 
the proton would stay stable since the Higgses $H_3$ and $\bar{H}_3$ 
do not match each other but get their masses rather by joining 
the states $\bar{H}'_3$ and $H'_3$ (in other words, 
$(\cM_3^{-1})_{11}$ would vanish). However, then the extra 
doublets $H'_2, \bar{H}'_2$ are rendered light and this 
would affect the gauge coupling unification.  


Thus, the proton decay rate crucially depends on $M_D$: 
\be{dim5-new}
(\cM_3^{-1})_{11} 
\sim \frac{M_D}{M_T^2} = 
\frac{M_D M_G}{M_T^2}\frac{1}{M_G}  
\ee
Therefore, the if $M_D$ is chosen so that $M_DM_G \leq M_T^2$,  
then the proton lifetime in this model would not exceed 
the one in the minimal $SU(5)$ model. 

Therefore, our theory below the GUT scale $M_G$ contains 
new states with masses $ \ll M_G$: 
two couples of triplets with masses $M_T,M'_T$ and 
a couple of doublets with mass $M_D$.  
Interestingly, if $M_T$ is about a mid geometrical between 
$M_D$ and $M_G$, then the presence of these extra states  
will not affect the fact of the gauge coupling unification 
and the value of the unification scale itself. 
The couple of doublets $H'_2+\bar{H}'_2$ contribute 
the gauge constants renormalization group running up 
from the scale $M_D$, while the two couples of triplets 
enter at the scale $M_T \sim \sqrt{M_DM_G}$. 
Therefore, at one loop approximation these extra states together 
contribute as a weakly split complete $SU(5)$ supermultiplet 
$5+\bar5$ living at the scale $M_D$.  

Therefore, for the given value of $\langle X\rangle$, 
one can always choose the enough large power $n$ for which the 
condition $M_DM_G \leq M_T^2$ is marginally satisfied. 
We obtain $\eps^{n+3}\leq \eps_G^3$, or in other words: 
\be{n-value} 
\frac{n+3}{3}\log \frac{M_P}{\langle X\rangle} > 
\log \frac{M_P}{\langle\Phi\rangle} , 
\ee 

On the other hand, now 50-plets are heavy and thus their impact 
on the gauge coupling running from the GUT to the string scale 
is not anymore so dramatic. 
Interestingly enough, in our model the MDM 
can be realized by using the standard Higgs 24-plet 
instead of 75-plet. Indeed, one can replace $\Phi$ 
in eq. (\ref{MDM-new})  as 
$\Phi \to \frac{1}{M_P} \Sigma \cdot \Sigma$, where 
$\Sigma \sim 24$ with a vanishing $U(1)_A$ charge has a VEV 
$\langle \Sigma\rangle \sim M_G$. (Hereafter such a model is 
refered as IMDM24 while the one involving the 75-plet as 
IMDM75.)  
This will result only in changing 
the order of magnitude of the triplet masses in (\ref{D-T}):  
$M_T,M'_T \sim \eps_G^4\eps^{-1}M_P\sim \eps^{-1}\cdot 10^{10}$ GeV. 
Then the condition  $M_DM_G \leq M_T^2$ translates into 
$\eps^{n+3}< \eps_G^7$. 

Implications of both IMDM75 and IMDM24 models for the extra 
states populating the big desert between the electroweak and GUT 
scales for different values of the scale $\langle X\rangle$ 
are shown in the illustrative Table 1. 
The values of the integer $n$ for which $M_T$ appears to be 
about a mid geometrical between $M_D$ and $M_G$.

\begin{table}
\caption{Masses of extra triplets and doublets in the IMDM models}
\label{t:tab1}
$$\begin{array}{|c|c|c|}
\hline & X=10^{17}~ {\rm GeV} & X=4\cdot 10^{17}~ {\rm GeV}   \\
& (\epsilon \simeq 1/20) & (\epsilon \simeq \sqrt{1/20}) \\
\hline {\rm IMDM75} & M_T= 4\cdot 10^{15}~ {\rm GeV} &
M_T= 10^{15} ~{\rm GeV}   \\ 
(\Phi \sim 75)& M_D= 2\cdot 10^{14}~{\rm GeV} &
M_D= 6\cdot 10^{13} ~{\rm GeV} \\ 
& (n=2) & (n=6) \\
\hline
{\rm IMDM24} & M_T= 4\cdot 10^{11}~ {\rm GeV} &
M_T= 10^{11}~ {\rm GeV}   \\ 
(\Sigma \sim 24)& M_D= 3\cdot 10^{6}~ {\rm GeV} &
M_D= 2\cdot 10^{5}~ {\rm GeV} \\ 
& (n=8) & (n=18) \\
\hline
\end{array}$$
\end{table}

\section{Neutrino masses } 

Until now 
we did not fix the value of $h$, since the needed pattern of 
the Higgs superpotential still allows the freedom for two 
uncorrelated $U(1)_A$ charges, say $Q_X$ and $Q_H$. 
However, the charge $h$ can be unambiguously fixed by 
considering the neutrino mass generating operators. 

The simplest assumption would be that the neutrino masses 
emerge from the Planck scale operators \cite{BEG}: 
\be{nu-MP} 
\frac{\beta_{ij}}{M_P} ~ \bar{5}_i \bar{5}_j H^2 
\ee 
which would be allowed by the $U(1)_A$ symmetry if 
$h=-\frac25 (n+3)q$. 
This operator induces the neutrino Majorana masses 
$m_\nu \sim M_W^2/M_P \sim 10^{-5}$ eV. 
This would suffice for explanation of the solar neutrino 
problem via the long wavelength ``just-so'' oscillation 
$\nu_e \to \nu_{\mu,\tau}$ provided that the 
mixing angles are large, of order 1. The latter would be the 
case if all (diagonal and non-diagonal) coupling 
constants $\beta_{ij}\sim 1$, unlike the Yukawa constants 
of the charged fermions. 
However, all other 
existing neutrino puzzles cannot be accomodated in this case.

In order to obtain the larger neutrino masses,  
one can use the seesaw mechanism \cite{seesaw}. 
Indeed, let us introduce the $SU(5)$ singlet neutrino states $N_m$ 
($m=1,2,\dots$),  which get the large Majorana masses 
via couplings to $X$ while their couplings to "left handed" 
neutrinos contained in $\bar5$ induce the Dirac mass terms. 
The relevant superotential terms can be taken as 
\be{seesaw} 
\Ga^D_{kl}H\bar5_k N_m  ~ + ~ \Ga^M_{lr} M_P N_l N_r 
\left(\frac{X}{M_P}\right)^{n'}
\ee 
which after integrating out the heavy $N$ states result in 
the following effective operators scaled by the inverse powers 
of $\langle X \rangle$:\footnote{  
The effective operators involving the direct powers of $X$, 
$\frac{X^{l}}{M_P^{l+1}} ~ \bar{5}^2 H^2$ 
would fix $h=-\frac15 (2n+6-l)q$.  
However, in this case the neutrino masses  
are too small and thus of no phenomenological interest. }   
\be{nu-op}
\frac{\beta_{ij}}{\eps^{n'}M_P} ~ \bar{5}_i \bar{5}_j H^2 
\ee
Therefore, for a given integer $n'=0,1,2,...$ 
($n'=0$ corresponds to case of operators (\ref{nu-MP})), 
these couplings fix the following relation between $h$ and $q$ charges: 
\be{h-nu}
h=-\frac15 (2n+n'+6)q
\ee 
Therefore, for $n'=4$ or $n'=5$ one obtains that 
$m_{\nu-\tau} \sim 1-10$ eV, 
which can provide the hot dark matter needed for the 
explanation of the large scale structure of the universe.  
If the neutrino masses obey the same hierarchy as the 
quark and lepton masses, e.g. 
$\nu_e: \nu_\mu : \nu_\tau \sim  u:c:t$, then the mass of 
$\nu_\mu$ will emerge in the range $\sim 3\cdot 10^{-3}$ eV 
which can explain the solar neutrino problem via the 
MSW mechanism. 

\section{Natural R parity } 

Thus, the charge $h$ can be fixed in terms of $q$ 
by considering the neutrino mass pattern. 
Now we show that for certain values of $h$ one can 
achieve a natural suppression of the $d=3$ and $d=4$  
B and L violating operators,  without imposing an {\it ad hoc} 
matter parity (R-parity). 

Indeed, if $h$ is given by eq. (\ref{h-nu}) then 
the $U(1)_A$ charge of the combinations $H\bar{5}_k$ equals to 
$-\frac12 n'q<0$. Therefore, these $d=3$ R parity breaking 
operators are forbidden at any power of $X/M_P$.  On the other hand, 
the charge of the combinations $10\cdot \bar5\cdot\bar5$ 
is $\frac12 (2n-n'+6)q$.  Hence, these operators could emerge only 
at the level 
\be{R-ops} 
10_i \bar5_j \bar5_k 
\left(\frac{X}{M_P}\right)^{(n+3)-n'/2}
\ee 
Thus, for an odd $n'$ no such operators can be built and the 
exact R-parity conservation emerges as an automatic (accidental) 
consequence of the $U(1)_A$ charge content of the fields in the 
theory. 

For an even $n'$ 
the R-parity breaking terms can emerge after substituting 
VEV of $X$, and thus they can be naturally suppressed, 
by the corresponding power of $\eps$.  
%
%
By taking for example  $n=2$ and $n'=2$, we obtain the 
following estimate for the constants of 
the R-parity breaking terms: $\la \sim \eps^4 
\sim 5\cdot 10^{-6}$ for $\eps\sim 1/20$. 
This is just at the border of the experimentally allowed 
region \cite{Vissani}  and can be of phenomenological 
interest for the testing in future experimens. 

In the models where the fermions carry the generation dependent 
$U(1)_A$ charges the R-parity breaking constants can be suppressed 
much stronger. For example, in the model of the section 7 
with the fermion charges as in eq. (\ref{10-5}) 
the following terms are allowed: 
\be{R-bre}
a_{ijk} 10_i \bar5_j \bar5_k \left(\frac{X}{M_P}\right)^{n+7-i} 
\ee 
Therefore, even for the constants $a_{ijk}\sim 1$, the 
R-parity breaking terms involving $10$ of the first family 
are suppressed as $\eps^8$ and thus they are well below the 
present experimental limits \cite{Vissani}.

\section{The Planck scale d=5 operators} 

In principle, in the minimal 
supersymmetric $SU(5)$ theory (in fact, already in the MSSM) 
the nucleon decay could be 
induced by the Planck scale $d=5$ operators \cite{Ellis}
$\sim \frac{\kappa}{M_P} (10_i 10_j)(10_l \bar5_k)$.\footnote{
The combinations in brackets can be in 5 or 45 channels, 
while the Higgsino mediated $d=5$ operators select only 
the channel 5.}  
The fact that these operators are cutoff by $M_P$ instead of 
$M_G$ as it takes place in the triplet Higgsino mediated case 
is no garancy that these Planck scale terms are safe unless 
the relevant coupling constants are further suppressed. 
For example, if $\kappa\sim 1$ then the operator 
$(q_2 q_1)(q_1 l_k)$ would induce the proton decay 
with $\tau(p\to K\nu)\sim \kappa^{-2} 10^{18}$ yr, 
i.e. much faster than the triplet Higgsino mediated $d=5$ 
operators do. 
Hence, in order to reconcile with the experimental bounds on 
the proton decay the constants $\kappa$ should be very small, 
$\kappa \leq 10^{-7}$. 


In our model these dangerous operators are 
indeed suppressed by the $U(1)_A$ symmetry. 
For the generation blind arrangement of the fermion charges 
(\ref{Q-f}) for the $U(1)_A$ charge of the relevant combination 
we have $Q(10^3\bar5)= (n+3)q$. Therefore, 
the lowest order term allowed is the following:  
\be{dim5-MP} 
\frac{1}{M_P} 10_i 10_j 10_l \bar5_k  
\left(\frac{X}{M_P}\right)^{n+3} 
~ \stackrel{X\to \langle X \rangle}{\Longrightarrow} 
~ \frac{\eps^{n+3}}{M_P} 10_i 10_j 10_l \bar5_k  
\ee
Then e.g. for $\eps\sim 1/20$ and $n=2$, which has been shown 
to be a reasonable choice for achieving 
the pattern for the Higgs superpotential parameters 
for achieving the proper suppression of the Higgsino mediated 
$d=5$ operators, we obtain $\kappa\sim 10^{-7}$ which is enough 
to reconcile the proton lifetime with the experimental limits. 

Anticipating the next section, the more suppression of 
the Planck scale $d=5$ operators can take place 
if fermions are 
prescribed to have the generation dependent $U(1)_A$ charges 
(see e.g. below, eq. (\ref{10-5})). In this case we would 
obtain  
\be{dim5-next}
\frac{1}{M_P} 10_i 10_j 10_l \bar5_k  
\left(\frac{X}{M_P}\right)^{n+14-i-j-l} 
\ee 
so that for $n=2$ the terms involving first and second generations 
of 10 are suppressed as $\eps^{12}$.

\section{ The U(1)$_A$ hierarchy of fermion masses } 

Below we try to join our voice to the common chorus \cite{IR} 
and attempt to incorporate the anomalous $U(1)_A$ symmetry 
for ``stitching the Yukawa quilt'' \cite{RRR} for understanding 
the quark and lepton mass and mixing pattern. 
We consider 
the $SU(5)$ IMDM where the Higgs superfields possess the $U(1)_A$ 
charges given as in eq. (\ref{Q-new}) while the fermion charges are 
flavour and generation dependent. 
Namely,  we choose the charges of 
$10_i=(u^c,q,e^c)_i$ and $\bar5_i=(d^c,l)_i$ as follows: 
\be{10-5} 
Q(10_i) = -\frac12 h + (3-i)q,  ~~~~ 
Q(\bar5_i) = \frac32 h + (n+5)q 
\ee 
where $i=1,2,3$ is a family index.  
Thus, only the top quark can get mass via the renormalizable 
Yukawa coupling $10_3 10_3 H$ while the masses of other fermions 
will emerge from the higher order operators including powers of 
$X/M_P$. The effective Yukawa superpotential 
including the higher order operators now has the form: 
\be{ij} 
W_{\rm Yuk}= 
f_{ij} 10_i 10_j H \left(\frac{X}{M_P}\right)^{6-i-j}  + 
g_{ik} 10_i \bar{5}_k \bar{H} \left(\frac{X}{M_P}\right)^{5-i} 
\ee 
with constants $\sim 1$ where $f_{ij}$ is symmetric, $f_{ij}=f_{ji}$, 
and $g_{ik}$ can be taken in the skew-diagonal form with vanishing 
elements $g_{13}$, $g_{23}$ and $g_{13}$ 
by a proper choice of the $\bar5$-plets basis.  

Therefore, after substituting the VEVs of the MSSM Higgs doublets 
$\langle H_u \rangle=v_u$ and $\langle H_d \rangle=v_d$ 
($v_u^2+v_d^2=v^2$, $v=174$ GeV, and $v_u/v_d=\tanb$). 
we would obtain the following pattern for the quark mass matrices: 
\be{Mu}
\begin{array}{ccc}
 & {\begin{array}{ccc} \,u^c_1 & \,\,~~ u^c_2 & \,\,~~ u^c_3
\end{array}}\\ \vspace{2mm}
\hat{m}^u= \begin{array}{c}
u_1 \\ u_2 \\ u_3  \end{array}\!\!\!\!\! &{\left(\begin{array}{ccc}
f_{11}\eps^4 & f_{12}\eps^3 & f_{13}\eps^2 \\ 
f_{12}\eps^3 & f_{22}\eps^2 & f_{23}\eps  \\ 
f_{13}\eps^2 & f_{23}\eps & f_{33} \end{array}\right)} 
\end{array}  \!\! \cdot v_u, ~~~~~
\begin{array}{ccc}
 & {\begin{array}{ccc} \, d^c_1 & \,\,~~ d^c_2 & \,\,~~ d^c_3
\end{array}}\\ \vspace{3mm}
\hat{m}^d= \begin{array}{c}
d_1 \\ d_2 \\ d_3  \end{array}\!\!\!\!\! &{\left(\begin{array}{ccc}
g_{11}\eps^4 & g_{12}\eps^4 & g_{13}\eps^4 \\ 
0 & g_{22}\eps^3 & g_{23}\eps^3  \\ 
0 & 0 & g_{33}\eps^2 \end{array}\right)} 
\end{array} \!\! \cdot v_d 
\ee 
where we take $\eps=\langle X\rangle/M_P\sim 1/20$, 
in accord to the estimate (\ref{xi}) provided that ${\rm Tr}Q$ 
has a moderate value.\footnote{Actually the value of ${\rm Tr}Q$ 
accumulated on the observable sector rather favours the 
larger value of $\eps$, say  $\eps \sim \sqrt{1/20}$. For this case  
one can simply modify the fermion charges so that expansion 
in the Yukawa superpotential (\ref{ij}) would go in powers of 
$(X/M_P)^2$. This will result only in change $\eps \to \eps^2$ 
in the fermion mass matrices and thus exactly the same pattern 
will be kept.} 
 Therefore, 
the horizontal hierarchy of quark masses exhibite the 
approximate scaling lows: 
\beqn{horiz} 
& m_t : m_c : m_u \sim 1 : \eps^2 : \eps^4  \nonumber \\   
& m_b : m_s : m_d \sim 1 : \eps : \eps^2  
\eeqn 
while for the CKM mixing angles we obtain the following estimates: 
\be{CKM} 
s_{12}, s_{23} \sim \eps, ~~~~~ s_{13}\sim \eps^2.   
\ee 
All these for $\eps\sim 1/20$ well agrees to the observed pattern 
of the quark masses and mixing 
(actually experimental value of the Cabibbo angle $s_{12}$ is 
about factor of 4 above this estimate, which 
can have an accidental origin).   

Note also that the vertical splitting between the quark masses 
also has a pattern which is favoured for $\tanb\sim 1$:  
\be{vert}
\frac{m_u}{m_d}\sim \tanb, ~~~~ 
\frac{m_c}{m_s}\sim \eps^{-1}\tanb, ~~~~ 
\frac{m_t}{m_b}\sim \eps^{-2}\tanb.  
\ee 
Recall, that small $\tanb$ is also favoured by the 
the proton lifetime.

The problem which remains is that since $X$ is a $SU(5)$ singlet, 
the second operator (\ref{ij}) leads to exactly the same 
mass matrix for the charged leptons as for the down quarks: 
$\hat{m}^e=\hat{m}^d$ in the $SU(5)$ limit, and thus 
$m_{e,\mu,\tau}=m_{d,s,b}$. 

The simplest possibility for evading these unwanted relations 
would be to involve the scalar $\Phi$ which breaks the $SU(5)$ 
symmetry into the quark and lepton mass generation. Then it 
woul induce different Clebsches for the down quark and lepton 
mass entries and thus remove these relations. 

However, we find that the simplest possibility is the 
following. Let us assume that the "Higgs" superfield $\bar{H}'$ 
carries the same charge as the fermion $\bar 5$-plets $\bar5_i$. 
According to eqs. (\ref{Q-new}) and (\ref{10-5}), this requires 
that $-h-q=\frac32 h + (n+5)q$ and thus $h$ is fixed as 
\be{h-model} 
h= -\frac25 (n+6)q 
\ee 
The $U(1)_A$ charges of all superfields involved in game in this 
case is given in Table 1. 

Then, if no {\em ad hoc} matter parity is introduced to 
distinguish the `fermion' $\bar5_i$ and the `Higgs' $\bar{H}'$  
superfields are essentially the same and 
thus $\bar{H}'$ can be identified as a fourth fermion $\bar5$-plet: 
$\bar{H}'=\bar5_4$. Then the second `Yukawa' term in (\ref{ij}) 
should be extended by including also $\bar{H}'$, so that the 
index $k$ now runs the values $k=1,2,3,4$.


\begin{table}
\caption{The $U(1)_A$ charges $Q$ of the superfields in units 
of $q=-Q_X$. The charges of the $SU(5)$ breaking fields 
$\Phi\sim 75$ or $\Sigma\sim 24$ in the above models are zero. }
\label{t:tab2}
$$\begin{array}{|c|ccccccccc|}
\hline & H & \bar H & \Psi & \bar{\Psi} & \Psi' & 
\bar{\Psi'} & H' &  \bar{H'},\bar {5}_i & 10_i \\
\hline Q  & -\frac{2(n+6)}{5} & -\frac{3(n+1)}{5}&  
 -\frac{2n+7}{5} & \frac{2(n+6)}{5} &
\frac{3(n+1)}{5} & -\frac{3n-2}{5} & 
\frac{3n-2}{5} & \frac{2n+7}{5} & \frac{n+21}{5} -i \\ 
\hline
\end{array}$$
\end{table}


On the other hand, all couplings presented for $\bar{H}'$ in the 
Higgs superpotential (\ref{W2-R}) and (\ref{W2-NR}) 
now are allowed also for $\bar5_k$, $k=1,2,3$.  
Then, without loss of generality, one can choose the basis in which 
the "fermions" $\bar5_i$ have no coupling to the 50-plet $\Psi$ 
in $W_2$ (\ref{W2-R}).  
In other words, $\bar{H}'=\bar{5}_4$ can be
defined as a combination of all $\bar5_k$ `fermions' 
which has the coupling $\bar{H}'\Phi\Psi$. 
However, in this basis the couplings 
\be{5-H} 
G_k\frac{X^{n+1}}{M_P^n} \bar5_k  H',  ~~~~ k=1,2,3,4  
\ee 
already cannot be rotated away and thus the doublet states 
in $\bar5_{1,2,3}$ cannot 
be identified anymore to the light lepton doublets $l_i$, 
but rather with the linear combinations of $\bar5_i$ and $\bar{H}'$ 
with the order 1 mixing angles. 

The `big' mass matrices of the down quark and lepton states now read
as: 
\be{Md-big}
\begin{array}{ccccc}
 & {\begin{array}{ccccc} 
\,\, d^c_1 & \,\,~~~ d^c_2 & \,\,~~~ d^c_3 & \,\,~~~ \bar{H}'_3 & 
\,\, ~~~ \bar{H}_3 
\end{array}}\\ \vspace{3mm}
\begin{array}{c}
d_1 \\ d_2 \\ d_3 \\ H'_3 \\ H_3 \end{array}\!\!\! 
&{\left(\begin{array}{ccccc}
m_{11}\eps^4 & m_{12}\eps^4 & m_{13}\eps^4 & m_{14}\eps^4 & 0 \\ 
0 & m_{22}\eps^3 & m_{23}\eps^3 & m_{24}\eps^3 & 0 \\  
0 & 0 & m_{33}\eps^2 & m_{34}\eps^2 & 0 \\ 
M^D_1 & M^D_2 & M^D_3 & M^D_4 & M'_T \\ 
0 & 0 & 0 & M_T & 0    
\end{array}\right)} \end{array}  
\ee 

\be{Ml-big}
\begin{array}{ccccc}
 & {\begin{array}{ccccc} 
\,\, e_1 & \,\,~~~ e_2 & \,\,~~~ e_3 & \,\,~~~ \bar{H}'_2 & \,~~~  \bar{H}_2 
\end{array}}\\ \vspace{3mm}
\begin{array}{c}
e^c_1 \\ e^c_2 \\ e^c_3 \\ H'_2 \\ H_2 \end{array}\!\!\! 
&{\left(\begin{array}{ccccc}
m_{11}\eps^4 & m_{12}\eps^4 & m_{13}\eps^4 & m_{14}\eps^4 & 0 \\ 
0 & m_{22}\eps^3 & m_{23}\eps^3 & m_{24}\eps^3 & 0 \\  
0 & 0 & m_{33}\eps^2 & m_{34}\eps^2 & 0 \\ 
M^D_1 & M^D_2 & M^D_3 & M^D_4 & 0 \\ 
0 & 0 & 0 & 0 & 0    
\end{array}\right)} \end{array}  
\ee 
where $M^D_k= G_k \eps^{n+2}M_P$ and $m_{ik}=g_{ik}v_d$ (in 
the upper $3\times 3$ block these terms can be still rotated 
to the skew diagonal form (\ref{Mu})).  

Now we see that the mixing of the  
states $d^c_{1,2,3}$ to the heavy triplets $\bar{H}_3,\bar{H}'_3$ 
will not affect significantly the upper $3\times 3$ block in 
the matrix (\ref{Md-big}), since 
the mixing angles are very small, $\sim M_D/M_T \ll 1$.  
Therefore, the down quark mass matrix remains practically the 
same as is given by eq. (\ref{Mu}). 

On the other hand, mixings of the states $e_{1,2,3}$ to 
$\bar{H}'_2=e_4$ are big.  
Hence, the doublet states 
in $\bar5_{1,2,3}$ cannot 
be identified anymore to the light lepton doublets $l_i$, 
but rather with the linear combinations of $\bar5_i$ and $\bar{H}'$ 
with the order 1 mixing angles. Clearly, this mixing will deviate 
the lepton mass matrix from the form given by eq. (\ref{Mu}) 
by the order 1 ``Clebsches'' in each element.

We gave up the matter parity. However, in the low energy 
theory (i.e. for the MSSM states) 
the R-parity breaking operators will be strongly 
suppressed. As we have already discussed in sect. 5, 
the $U(1)_A$ charge of the combinations $H\bar5_k$ is always 
negative, $-q$, and thus this terms cannot emerge at any 
order of $X/M_P$. As for the combination 
$10_i \bar5_j \bar5_k$, its $U(1)_A$ charge is positive, 
$n+7-i$ (Note, it depends only on the generation index of 
10, $i=1,2,3$). Therefore, these terms can emerge in the form 
(\ref{R-bre}) written down in sect. 5. However it was remarked 
that the R-parity breaking constants are much below the 
experimental bounds. 

Below we present some other interesting R-parity breaking terms 
allowed in our model. The ones involving only the light 
(MSSM) states are: 
\be{R-break}
\frac{1}{M_P}
10_i 10_j 10_k \bar{H} \left(\frac{X}{M_P}\right)^{12-i-j-k} 
\ee
with an impact for the lightest  neutralino stability. 
All other possible operators involve the superheavy 
states. E.g. the coupling 
\be{HH'10}
10_i \bar{H}\bar{H}' \left(\frac{X}{M_P}\right)^{5-i}
\ee 
which can even faster destabilize the lightest neutralino, via the 
diagram induced by the exchange of the colour triplet Higgsino 
with a mass $M_T\sim \langle\Phi\rangle^2/\langle X\rangle
\sim 10^{15}$ GeV: $n\to udd$ or $n\to d u^c e^c$. 
Therefore, if these processes are active 
in this model, the cold dark matter of the universe cannot consist 
of the lightest neutralino and another candidate should be found.   
 

 

\section{Discussion } 

In this paper we have revised the missing doublet $SU(5)$ model. 
A key role in our picture is played by the anomalous gauge 
symmetry $U(1)_A$.  
We have shown some examples of supersymmetric 
$SU(5)\times U(1)_A$ models which could 
provide an appealing simoultaneous solution to the 
various ``hot'' puzzles in the SUSY GUT philosophy as are 
the gauge hierarchy and doublet-triplet splitting problem, 
problem of fermion mass hierarchy, origin of matter parity 
(or R parity) conservation and so long lifetime of proton.  
In particular, the $U(1)_A$ charge content of superfields 
in the theory can be arranged so that R parity breaking 
operators will be forbidden at any order in $M_P^{-1}$. 
In other words, the exact conservation of R parity  
can be an accidental consequence of the gauge symmetry. 
In more interesting cases, however, the R parity can 
be only approximate symmetry, with implications for 
the future search of the R-parity breaking phenomena and 
certainly with a great impact on the stability of the 
lightest neutralino.

We have also extended a picture for the fermion masses 
by involving $U(1)_A$ as a horizontal symmetry. 
The fermion mass hierarchy as well as the magnitudes 
of the CKM mixing angles can be naturally understood in 
terms of small parameter ($\eps\sim 1/20$) or not so small 
($\eps\sim \sqrt{1/20}$), with a proper choice of the fermion 
$U(1)_A$ charges. We have also shown a simple mechanism 
for explaining the origin of the about factor of 3 splitting 
between the down quark and lepton masses in the same family. 
Interestingly, this mechanism is organically related to the features 
needed for achieving the realistic IMDM patter for the 
doublet-triplet splitting. 

The actual value of the $U(1)_A$ symmetry breaking scale cannot 
be deduced unambiguoulsy. The Green-Schwarz mechanism implies 
the following relations between the mixed $U(1)_A$ anomaly 
coefficients and the Kac-Moody levels:  
\be{anom}
\frac{C_5}{k_5} = \frac{C_A}{k_A} = \frac{C_g}{k_g} 
\ee 
In our models all anomaly coefficients are typically $O(100)$. 
Among these the mixed $U(1)_A$ -- $SU(5)$ anomaly $C_5$ 
is contributed solely by the $SU(5)$ superfields in the game 
unless some other states are also introduced. 
The value ${\rm Tr}Q/q$, which is essentially the mixed 
$U(1)_A$-gravity anomaly $C_g$ remains arbitrary, since besides 
the observable ($SU(5)$) matter it can be contributed by 
some `hidden' matter singlet with respect to $SU(5)$: 
${\rm Tr}Q={\rm Tr}Q_{\rm obs} + {\rm Tr}Q_{\rm hid}$. 
These additional states will contribute also the 
$U(1)_A^3$ anomaly coefficient $C_A\sim {\rm Tr}(Q^3)$. 
In the IMDM models the `observable' portion of $C_g$ is 
rather large, ${\rm Tr}Q_{\rm obs}\sim 100$, and it would favour 
the `not so small' value of $\eps$ (say $\sim \sqrt{1/20}$).  
If the hidden matter gives a big negative contribution, 
then $\eps\sim 1/20$ can be also possible. 
There can be also a case that the field $X$ shares the VEV 
(\ref{xi}) with some other singlets so that its its own portion 
in $\sqrt{xi}$ is small. 
In any case, the Diophantian equations to be satisfied 
for having the Green-Schwarz mechanism valid for the 
universal cancellations of all mixed anomalies through the shift of 
the axion ${\rm Im}s$ do not restrict much the 
charge content in our scheme 
unlike the cases of the MSSM based models \cite{IR},
since now the MSSM is embedded in the $SU(5)$. 
In particular, the GUT scale value of the Weinberg angle 
$\sin^2\Theta_W=3/8$ now directly follows from $SU(5)$ 
without specific matching of the mixed anomalies 
for each factor in $SU(3)\times SU(2)\times U(1)$ separately.  

We have not discussed neither the origin 
of the supersymmetry breaking nor the $\mu$-problem.  
It was implicitly assumed that the supersymmetry 
breaking occurs via the incognito mechanism in some 
hidden sector and then transmitted through the gravity 
to the observable sector.  Perhaps there 
can be found a clever mechanism which relates the origin of 
the supersummetry breaking to the same anomalous $U(1)_A$ 
symmetry, in the spirit recently discussed in refs. 
\cite{gia-alex}. We have tried some not so clever possibilities 
but did not find a one worth of publishing. 

The interesting features of the IMDM models 
make us to think that maybe neither the missing 
doublet mechanism nor the concept of the anomalous $U(1)_A$ 
stringy gauge symmetry are not that bad ideas if the two are 
working together. 

Of course, there is a big question whether the models presented 
here or alike can indeed emerge from the string theory. 
No explicit string construction exists at the Kac-Moody level 
$k_5\geq 4$ which is required for having 50-plets in the 
$SU(5)$ stringy GUT \cite{Dienes}. However, if the 
the IMDM like models can be indeed found in a string theory 
context, they would have the interesting feature that 
the string constant 
$g_{\rm str}^2=k_5g_5^2=k_Ag_A^2$ could be enough large, 
say $\sim 5$.  As we already remarked, in the IMDM models 
(perhaps more interesting is IMDM24), 
the $SU(5)$ gauge constant running up from the GUT scale 
can be prevented from being `exploded' to the strong coupling 
regime below $M_P$. However, its value at energies $\sim M_P$ 
should be significantly larger than in the minimal $SU(5)$, 
due to contributions of the 50-plets and 75- or 24-plet. 
Therefore, it is rather probable that $\al_{\rm str}$
can have a magnitude of about 0.5-1, which can leave open the 
possibility to prevent the runaway behaviour and find the 
true vacuum values for the dilaton and the moduli fields. 
This feature might be of certain interest since 
by the duality arguments neither 
the weak regime $\al_{\rm str}\ll 1$ nor the strong one 
$\al_{\rm str}\gg 1$ seem to be good for stabilizing the 
vacuum state \cite{Witten}.

\section{Acknowledgements} 

We thank Gia Dvali for sending to us the preprint of ref. \cite{Gia}.

\end{document}